\documentclass[twocolumn]{IEEEtran}
\usepackage{cite}
\usepackage{amsmath,amsthm,amssymb,amsfonts}
\usepackage{graphicx}
\usepackage{textcomp}
\usepackage{xcolor}
\usepackage{color}
\usepackage{setspace}

\usepackage{mathrsfs}

\usepackage{bm}
\usepackage{pgfplots}
\usepackage{tikz}
\usetikzlibrary{arrows}
\usepackage{subfigure}
\usepackage{graphicx,booktabs,multirow}
\usepackage{upgreek}
\usepackage{bm}
\usepackage{mathrsfs}
\usepackage{algorithm}
\usepackage{algorithmicx}
\usepackage{algpseudocode}

\definecolor{colorhkust}{RGB}{20,43,140}
\definecolor{colortsinghua}{RGB}{116,52,129}
\definecolor{color1}{RGB}{128,0,0}

\newtheorem{lem}{Lemma}

\mathchardef\re="023C
\mathchardef\im="023D
\IEEEoverridecommandlockouts
\begin{document}

\title{Joint  Activity Detection and Channel Estimation in Massive Machine-Type
Communications with Low-Resolution ADC}

\author{
   \IEEEauthorblockN{Ye Xue$^{1}$, An Liu$^{2}$, Yang Li$^{1}$, Qingjiang Shi$^{1,3}$, and Vincent Lau$^{4}$}\\
  \IEEEauthorblockA{$^{1}$ Shenzhen Research Institute of Big Data, Shenzhen 518172, China\\
  	$^{2}$ College
of Information Science and Electronic Engineering, Zhejiang University,
Hangzhou 310027, China\\
$^{3}$ Tongji University, Shanghai 201804, China\\
$^{4}$ Dept. of ECE, The Hong Kong University of Science and Technology, Hong Kong\\
  }
  \thanks{The work of Y. Xue was supported by the National Key R$\&$D Program of China under grant 2022YFA1003900. The work of Q. Shi was supported  by the National Natural Science Foundation of China (NSFC) under grant 62231019. The work of Y. Li was supported  by the  NSFC under grant 62101349.}

}

\maketitle

\begin{abstract}
In massive machine-type communications, data transmission is
usually considered sporadic, and thus inherently has a sparse structure. This paper focuses on the joint  activity detection
(AD) and channel estimation (CE) problems in  massive-connected
 communication systems with low-resolution
analog-to-digital converters. To further exploit the sparse structure in transmission, we propose a maximum posterior probability (MAP) estimation problem based on both  sporadic activity and sparse channels for  joint AD and CE. Moreover, a majorization-minimization-based method is proposed for solving
the MAP problem. Finally, various numerical experiments verify that the proposed scheme outperforms   state-of-the-art
methods.
\end{abstract}

\section{Introduction}

Future wireless networks will face a dramatically increasing density
of devices, especially in  massive machine-type communications (mMTC).   This makes accurate
 activity detection (AD) and channel estimation (CE) very challenging due to the high-dimensional channel matrices and the limited number of pilots.

Fortunately,  the sporadic transmission  property  \cite{boccardi2014five} enables  sparse recovery methods for  AD and  CE problems
in mMTC \cite{xu2015active,hannak2015joint,liu2018massive,chen2018sparse,8700603,9606865,9028221}.
Specifically, \cite{xu2015active} explored the sparse structure
of the active devices and proposed a modified Bayesian compressive sensing
 algorithm to perform joint AD and CE in cloud radio access
networks. References \cite{hannak2015joint,liu2018massive,chen2018sparse}
resorted to the  message-passing-based method to estimate
the sparse channel matrices and then detected the active devices by comparing the estimated channel power of each device in massive-connected systems. 
However, these works have the full-resolution assumption of the signal which  prevents them from being deployed in practical systems with impairments. References \cite{8700603,9606865,9028221} considered low-resolution analog-to-digital converters (ADC) in the system but they  only exploited the sparsity from the
sporadic activity, ignoring the sparsity in the angular domain channel. However, by further exploiting the angular domain channel sparsity, it would be possible to improve the AD and CE performance.

In this paper, we jointly exploit the sparsity in the device activity and the angular domain channel  to address the joint AD and CE problem in mMTC systems with low-resolution ADC. The main contributions
are summarized below.
\begin{itemize}
\item \textbf{Hierarchical Joint Sparsity of Device Activity and Channel:}
We consider joint group sparsity introduced by the  sporadic device activity
and the massive MIMO channel with a hierarchical sparsity structure\footnote{An inactive device will lead to the corresponding column of the channel
matrix being zero. On the other hand, the channel of an active device
will exhibit burst sparsity induced by the scattering cluster \cite{tse2005fundamentals},
as illustrated in Fig. \ref{fig:-Uplink-system}.}. To fully explore this structured sparsity, we propose a  two-level hierarchical sparsity model. Such a model
can accurately represent the joint sparsity as well as enable a tractable solution. 
\item \textbf{Maximum posterior probability (MAP)-based Problem Formulation with Bussgang Decomposition:} To solve the joint AD and CE problem, we start from the
MAP estimation criterion. However, naive MAP formulation will lead to a  non-elementary likelihood. To address this issue,  we propose a likelihood  based on
the Bussgang decomposition \cite{bussgang1952crosscorrelation,li2017channel,jacobsson2017quantized}. The resultant likelihood is  quadratic, which dramatically releases the computational burden.

\item \textbf{Majorization-Minimization (MM)-based Algorithm
Design:}  We then propose
to solve the MAP problem by MM-based method with a novel surrogate design. In each
iteration, the proposed algorithm can be reduced to solving a quadratic 
sub-problem with the second-order coefficient matrix being diagonal. 
\end{itemize}
\emph{Notations}: $\bm{X}^{-1},\bm{\thinspace X}^{T}$, $\bm{X}^{*}$
and $\bm{X}^{H}$ denote the inverse, transpose,
conjugate, and conjugate transpose of the matrix $\bm{X}$,
respectively. $Re(\bm{X})$ is the real-value decomposition of the complex matrix
$\bm{X}$. $Vec(\bm{X})$ is the vectorization of  matrix $\bm{X}$. $diag([a_1,\ldots,a_N]^T)$ is the diagonal matrix with $a_1,\ldots,a_N$ being the diagonal elements. The notation $\otimes$ is the Kronecker product. We use $\mathcal{CN}(x;\mu,\nu)$
and $\mathcal{N}(x;\mu,\nu)$ to denote the PDF of a complex Gaussian
and a real Gaussian random variable $x$ with
mean $\mu$ and variance $\nu$, respectively. The notation $\bm{R}_{\bm{x}\bm{y}}$ is the correlation matrix of random vectors $\bm{x}$ and $\bm{y}$.

\section{System Model\label{sec:System-Model}}

\subsection{Uplink Signal Model}\label{subsec:Uplink-Random-Access}

\begin{figure}[htbp]
\begin{centering}
\includegraphics[scale=0.65]{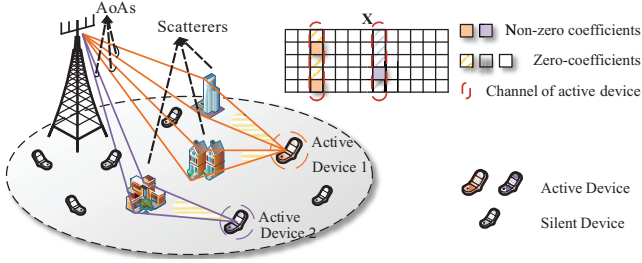}
\par\end{centering}
\caption{ Uplink system with both sparse device activities and sparse angular domain
channels\label{fig:-Uplink-system}.}
\end{figure}

In this work, we consider a single-cell uplink system
with an $M$-antenna base station (BS) ($M\gg1$) simultaneously serving $N$ single-antenna devices. These $N$ devices  exhibit  sporadic activity as shown in Fig.\ref{fig:-Uplink-system}.  

The activity of the $n$-th device in a particular time slot is given as
\begin{equation}
s_{n}=\begin{cases}
1, & \text{if the \ensuremath{n}-th device is active,}\\
0, & \text{otherwise}.
\end{cases}\label{eq:Eqin}
\end{equation}When the $n$-th device is active, it transmits a pilot sequence with
$T$ symbols. Hence, the received signal $\bm{Y}\in\mathbb{C}^{M\times T}$
at the BS is given by
\begin{align}
\bm{Y} & =\Sigma_{n=1}^{N}s_{n}\sqrt{g_{n}}\bm{h}_{n}\bm{d}_{n}^{T}+\bm{V}\label{eq:Eqs1}\\
 & =\bm{H}\bm{S}\bm{G}^{1/2}\bm{D}+\bm{V}.\nonumber 
\end{align}
\\
{In Eq.(\ref{eq:Eqs1}), $\bm{h}_{n}\in\mathbb{C}^{M\times1}$ represents the quasi-static small-scale fading channel between the $n$-th device and the BS, and $\sqrt{g_{n}}$ is the effective power factor combining the path loss and transmit power of the $n$-th device, which is assumed to be known at the receiver. $\bm{d}_{n}=[d_{n,1},\dots,d_{n,T}]\in\mathbb{C}^{T\times1}$ is the pilot signal assigned to the $n$-th device. $\bm{V}\in\mathbb{C}^{M\times T}$ is the complex Gaussian noise matrix with zero mean and covariance
matrix $\sigma_{v}^{2}\bm{I}$. We also have $\bm{H}=[\bm{h}_{1},\ldots\bm{h}_{N}]\in\mathbb{C}^{M\times N}$,
$\bm{S}=diag([s_{1},\ldots,s_{N}]^T)$, $\bm{G}^{1/2}=diag([\sqrt{g_{1}},\ldots,\sqrt{g_{N}}]^T)$
and $\bm{D}=[\bm{d}_{1},\ldots,\bm{d}_{N}]^T\in\mathbb{C}^{N\times T}$.
}

\subsection{Channel Sparsity in Angular Domain }

{We consider flat-fading massive MIMO channels with
limited scatterers around the BS. Applying the widely used discrete multipath channel model in \cite{tse2005fundamentals}, the uplink
channel response from the $n$-th device to the BS can be modeled as}

{
\begin{equation}
\bm{h}_{n}=\sum_{c=1}^{N_{c}}\zeta_{c,n}\bm{u}\left(\theta^{c}\right),\label{eq:AH-1}
\end{equation}
where $N_{c}$ is the number of scattering clusters, $\zeta_{c,n}$ stands for the complex channel gain corresponding to the $c$-th scattering cluster path of the $n$-th device, $\theta^{c}$ represents the azimuth angle of arrival (AoA) corresponding to the $c$-th scattering
cluster, and $\bm{u}\left(\theta^{c}\right)\in\mathbb{C}^{M}$ is the steering vector for the BS antenna array. The number of the scattering clusters $N_{c}$ is usually much smaller than the number of antennas $M$ in the massive MIMO regime. Specifically, for a  half-wavelength space uniform linear array (ULA), the steering vector has the form
\begin{align}
 & \bm{u}\left(\theta\right)=\nonumber \\
 & \frac{1}{\sqrt{M}}\left[1,e^{-j\pi\sin\left(\theta\right)},e^{-j2\pi\sin\left(\theta\right)},\cdots,e^{-j(M-1)\pi\sin\left(\theta\right)}\right]^{T}.\label{eq:arrayu}
\end{align}
Let $\{\theta_{1},\ldots,\theta_{\tilde{M}}\}$ be a uniform sampling grid, which covers the angular spread $[-\frac{\pi}{2},\frac{\pi}{2}]$.
We then define $\bm{U}_{R}=\left[\bm{u}\left(\theta_{1}\right),...,\bm{u}\left(\theta_{\tilde{M}}\right)\right]\in\mathbb{C}^{M\times\tilde{M}}$
and $\bm{\bar{H}\in}\mathbb{C}^{\tilde{M}\times N}$ as the array response matrix and the angular domain channel matrix, respectively.
Then, $\bm{H}$ can be expressed in a compact form as
\begin{equation}
\bm{H}=\bm{U}_{R}\bm{\bar{H}}.\label{eq:AHcomp}
\end{equation}
Note that the angular domain channel matrix $\bm{\bar{H}\in}\mathbb{C}^{\tilde{M}\times N}$ is, in general, sparse due to the limited local scattering effects at the BS side in massive MIMO systems\cite{zhou2007experimental}.

{Let $\bm{\text{\ensuremath{\bm{X}}}=\bar{H}S}$ being the {\em aggregated variable}, 
the uplink signal model (\ref{eq:Eqs1}) can be expressed as
\begin{align}
\bm{Y} & =\bm{U}_{R}\bm{\bar{H}}\bm{S}\bm{G}^{1/2}\bm{D}+{\bm{V}}\label{eq:sys2}\\
 & =\bm{U}_{R}\bm{X}\bm{G}^{1/2}\bm{D}+{\bm{V}}.\nonumber 
\end{align}
}

\subsection{Receiver with Low-Resolution ADC\label{subsec:Receiver-with-Few-Bit}}

At the BS, we consider the low-resolution ADC effect and focus on the scenario where the thermal noise
in the system appears before the signal is quantized. Furthermore,
we assume the received signal $\bm{Y}$ is quantized component-wisely
and separately to the real and imaginary parts with a general $B$-bit
scalar quantizer (not necessarily the uniform quantizer). The signal
after quantization is given by\\
\begin{equation}
\bm{R}=Q_{B}^{c}(\bm{Y}),\label{eq:quantr}
\end{equation}
where $Q_{B}^{c}(\cdot)$ is applied component-wise. Specifically,
for each component, we have 
\begin{equation}
Q_{B}^{c}(Y_{(m,n)})=Q_{B}(Re\{Y_{(m,n)}\})+jQ_{B}(Im\{Y_{(m,n)}\}).
\end{equation}
Each real-valued $B$-bit scalar quantizer, $Q_{B}$, maps the real-valued
input onto a finite-cardinal quantization alphabet $\{\beta_{l_{B}}:\beta_1,\ldots,\beta_{2^{B}}\}$ according to $2^{B}-1$
quantization thresholds: $\{\alpha_{l_{B}}:\ensuremath{-\infty<\alpha_{1}<\alpha_{2}<,\ldots,<\alpha_{2^{B}-1}<\infty}\}$.
We define $\alpha_{0}=-\infty$ and $\alpha_{2^{B}}=\infty$ for notation
convenience. The output of each real-valued scalar quantizer is $\beta_{l_{B}}$
if the input belongs to the interval $(\alpha_{l_{B}-1},\alpha_{l_{B}}]$.
We further define $\Delta_{l_{B}}=$$\beta_{l_{B}}-\beta_{l_{B-1}}$
as the quantization step size.

\section{MAP-based Problem Formulation
\label{sec:MAP}}

The joint AD and CE is modeled as an MAP estimation problem with respect
to the aggregated variable $\bm{X}$ in this section. We shall first elaborate
on the hierarchical prior probability model on $\bm{X}$. Then we will present the likelihood approximation via Bussgang decomposition. Based
on these two components, we will elaborate on the MAP formulation.

\subsection{Hierarchical Prior Probability of Aggregate Sparse Variable \label{subsec:Hierarchical-prior-information} }

The aggregated variable $\text{\ensuremath{\bm{X}}}=\bm{\bar{H}S}$
has two sources of sparsity, namely, the sparsity induced by the sporadic
device activity $\bm{S}$ and the group sparsity induced by the
limited scattering in $\bm{\bar{H}}$. We will illustrate the
prior of $\bm{S}$ and $\bm{\bar{H}}$ separately and then
give the overall two-level hierarchical prior for $\bm{X}$.

\subsubsection{Prior for \textcolor{black}{device activity} $\bm{S}$}

As illustrated in Section \ref{sec:System-Model}, $\bm{S}=diag([s_{1},\ldots,s_{N}]^T)$
represents the device activity with i.i.d. binary elements. Assume that
we have  knowledge of the device active ratio $q_{s}$. Then, the
prior probability of the activity of the $n$-th MTC device is given
by

\begin{equation}
p(s_{n})=q_{s}^{s_{n}}(1-q_{s})^{1-s_{n}}.\label{eq:priors-1}
\end{equation}

\subsubsection{Prior for the angular domain channel $\bm{\bar{H}}$}

Following the commonly used sparse Bayesian model \cite{tipping2001sparse},
we assign a Gaussian prior distribution with distinct precision for
each element of $\bm{\bar{H}}$. The prior of $\bar{H}_{(m,n)}$
is given by

\begin{equation}
p(\bar{H}_{(m,n)}|\gamma_{(m,n)})=\mathcal{CN}(\bar{H}_{(m,n)};0,\gamma_{(m,n)}^{-1}).\label{eq:phgamma}
\end{equation}
The precision $\{\gamma_{(m,n)}:m=1,\ldots,M,n=1,\ldots N\}$ is given
by i.i.d. Gamma distribution

\begin{equation}
p(\gamma_{(m,n)})=\Gamma(\gamma_{(m,n)};a,b)\text{,}\label{eq:gamma}
\end{equation}
where $a$, $b$ are the hyperparameters. With an appropriate
choice of the hyperparameters $a$ and $b$,\textcolor{blue}{}\footnote{In this paper, we set $a$,$b\to0$ as in \cite{DAI2017FDD} 
to obtain a broad hyperprior.}\textcolor{blue}{{} }the distribution $p(\bar{H}_{(m,n)})$ is recognized
as encouraging sparsity due to the heavy tails and sharp peak at zero
\cite{ji2008bayesian}. The overall prior distribution of $\bar{H}_{(m,n)}$
is given by

\begin{align}
p(\bar{H}_{(m,n)}) & =\int_{-\infty}^{\infty}p(\bar{H}_{(m,n)}|\gamma_{(m,n)})p(\gamma_{(m,n)})d\gamma_{(m,n)}\label{eq:ph-1}\\
 & =\frac{a}{\pi b}\left(\frac{||\bar{H}_{(m,n)}||^{2}}{b}+1\right)^{-(a+1)}.\nonumber 
\end{align}
\\

\subsubsection{Hierarchical prior for the aggregated variable
$\bm{X}$}

According to (\ref{eq:priors-1})\textendash (\ref{eq:ph-1}) and
the relation $\text{\ensuremath{\bm{X}}}=\bm{\bar{H}S}$,
we have 

\begin{align}
p(\bm{X})= & \prod_{n=1}^{N}\Bigg(\bigg(p(s_{n}=1)\prod_{m=1}^{M}p(X_{(m,n\text{)}}|s_{n}=1\big)\bigg)\label{eq:px}\\
 & +\bigg(p(s_{n}=0)\prod_{m=1}^{M}p(X_{(m,n)}|s_{n}=0)\bigg)\Bigg).\nonumber 
\end{align}
In Eq.  (\ref{eq:px}), we have $p(X_{(m,n)}|s_{n}=1)=p(\bar{H}_{(m,n)})$
and $p(X_{(m,n)}|s_{n}=0)=\delta(X_{(m,n)})$. To avoid the undesirable singularity of  $\delta(\cdot)$, we consider the following approximation of $\delta(X_{(m,n)})$:

\begin{equation}
p(X_{(m,n)}|s_{n}=0)=\mathcal{CN}(X_{(m,n)};0,\epsilon),
\end{equation}
where $\epsilon$ is a very small value.

As such, the logarithmic prior of $\bm{X}$ is given by
\begin{align}
&\log p(\bm{X})  =\log p(\bm{x})=\sum_{n=1}^{N}\log\Bigg[\sum_{s_{n}\in\{0,1\}}\Bigg(\bigg((1-q_{s})\label{eq:prior}\\
  \times&\frac{1}{(\pi\epsilon)^{M}}\prod_{i_{r}=(n-1)M+1}^{nM}\exp(-\frac{x_{i_{r}}^{2}+x_{i_{r}+MN}^{2}}{\epsilon})\bigg)^{1-s_{n}}\nonumber \\
\times & \bigg(q_{s}(\frac{a}{\pi b})^{M}\prod_{i_{r}=(n-1)M+1}^{nM}(\frac{x_{i_{r}}^{2}+x_{i_{r}+MN}^{2}}{b}+1)^{-(a+1)}\bigg)^{s_{n}}\Bigg)\Bigg]\nonumber 
\end{align}
where $\bm{x} \triangleq Re(Vec(\bm{X}))$.

\subsection{\label{subsec:Likelihood-representation-via}Approximation of the Likelihood Function via Bussgang Decomposition}

Another important component in the MAP formulation is the likelihood function $p(\bm{R}|\bm{X})$. Due to the quantization, writing $p(\bm{R}|\bm{X})$ in a direct way \cite{mo2018channel,liu2019generalized,wen2016bayes} will result in a very complicated and highly non-linear function, which will lead to difficult optimization and numerical issues. To address these drawbacks, we adopt the Bussgang decomposition \cite{bussgang1952crosscorrelation} to model the quantization effect in the derivation of the likelihood function. By the Bussgang decomposition, the likelihood can be approximated by a Gaussian likelihood, which leads to a more tractable optimization problem. 

For mathematical convenience, we first reformat $\bm{Y}$ in
Eq.  (\ref{eq:sys2}) into a real-valued vector. The real-valued  receive signal after vectorization  is given by
\begin{align}
\bm{r} & =Q_B^c(\bm{y})=Q_B^c(\tilde{\boldsymbol{\Phi}}\bm{x}+\bm{v}), 
\end{align}
where $\bm{y}=Re(Vec(\bm{Y}))\in\mathbb{R}^{2MT}$, $\tilde{\boldsymbol{\Phi}}= Re(Vec(\bm{G}^{1/2}\bm{D})^{T}\otimes\bm{U}_R))\in\mathbb{R}^{2MT\times2MN}$, $\bm{x}=Re(Vec(\bm{X}))\in\mathbb{R}^{2MN}$ and $\bm{v}=Re(Vec(\bm{V}))\in\mathbb{R}^{2MT}$.
Then, using Bussgang decomposition\cite{bussgang1952crosscorrelation}, the received signal after quantization can be expressed as
\begin{equation}
\bm{r}\cong\bm{K}\tilde{\boldsymbol{\Phi}}\bm{x}+\bm{K}\bm{v}+\bm{n}_{q}=\bm{A}\bm{x}+\bm{z},\label{eq:expressr}
\end{equation}
where $\cong$ means equivalent upto the second-order statistics, $\bm{K} =\bm{R}_{ry}\bm{R}_{yy}^{-1}$, $\bm{A}=\bm{K}\tilde{\boldsymbol{\Phi}}$, and $\bm{z}=\bm{K}\bm{v}+\bm{n}_{q}$ is the effective noise. The correlation matrix of the  residual noise $\bm{n}_{q}$ in Bussgang decomposition  is 
$\bm{R}_{\bm{n}_{q}\bm{n}_{q}}=\bm{R_{rr}}-\bm{K}\bm{R}_{yy}\bm{K}^{H}$. 
Therefore, the covariance matrix of $\bm{z}$ is given by
\begin{equation}
\boldsymbol{\Sigma}=\frac{\sigma^2_v}{2}\bm{KK}^{T}+\bm{R}_{\bm{n}_{q}\bm{n}_{q}}.\label{eq:sigexp}
\end{equation}
 Then, the log-likelihood function is  then given by\footnote{ Since Gaussian
noise is the worst-case noise with respect to reducing information
\cite{li2017channel,hassibi2003much,diggavi2001worst}, we can approximate
the likelihood function $p(\bm{r}|\bm{x})$ by assuming $\bm{n}_{q}$
is Gaussian. }
\begin{align}
\log p(\bm{r}|\bm{x}) & =-\frac{1}{2}(\bm{r}-\bm{A}\bm{x})^{T}\boldsymbol{\Sigma}^{-1}(\bm{r}-\bm{A}\bm{x})\label{eq:lik}\\
 & -\log\sqrt{(2\pi)^{2MT}\det(\boldsymbol{\Sigma})}\nonumber 
\end{align}
Combining the results of the log prior distribution $p(\bm{x})$ in (\ref{eq:prior}) and
the log-likelihood $p(\bm{r}|\bm{x})$ in (\ref{eq:lik}), the MAP problem is formulated
as
\begin{align}
\underset{\bm{x}\in\mathbb{R}^{2NM}}{\text{minimize}}\quad & f(\bm{x}),\label{eq:prob1}
\end{align}
where $f(\bm{x})=-\log p(\bm{r}|\bm{x})-\log p(\bm{x})\propto-\log p(\bm{x}|\bm{r})$.

\section{MM-based Algorithm for Joint AD and CE 
\label{sec:Proposed-Joint-Activity}}

\subsection{MM-based Estimation of the Aggregated Variable}
In this section, we adopt an MM-based method to iteratively
find a stationary point of the non-convex problem (\ref{eq:prob1}).
The MM-based algorithm requires an effective surrogate function.
To design the surrogate function, we need the following lemma:
\begin{lem}
{[}Majorization of a Quadratic Form{]} \label{lem:major2}Let $\boldsymbol{\Omega}\in\mathbb{R}^{L\times L}$
be a symmetric positive semi-definite  matrix. Then the quadratic
form $\bm{x}^{T}\boldsymbol{\Omega}\bm{x}$ is majorized by
$\bm{x}^{T}\tilde{\boldsymbol{\Omega}}\bm{x}-2\bm{x}^{T}(\tilde{\boldsymbol{\Omega}}-\boldsymbol{\Omega})\bm{x}_{0}+\bm{x}_{0}^{T}(\tilde{\boldsymbol{\Omega}}-\boldsymbol{\Omega})\bm{x}_{0}$,
with $\tilde{\boldsymbol{\Omega}}\succeq\boldsymbol{\Omega}$, where
$\bm{x}_{0}\in\mathbb{R}^{L}$ is any fixed point.
\end{lem}
\begin{IEEEproof}
$(\bm{x}-\bm{x}_{0})^{T}(\tilde{\boldsymbol{\Omega}}-\boldsymbol{\Omega})(\bm{x}-\bm{x}_{0})\geq0,\forall\bm{x}$
given $\bm{x}_{0}$.
\end{IEEEproof}
We apply the expectation\textendash maximization procedure and
the Taylor expansion of $-\text{log}P(\bm{x})$ to deal
with the non-convexity issue, and apply Lemma \ref{lem:major2} to
simplify the surrogate function. At the $j$-th iteration, the surrogate function is given by

\textcolor{black}{\small{}
\begin{align}
 & \hat{f}(\bm{x}|\bm{x}^{(j)})\label{eq:surr1}\\
= & \frac{1}{2}\bm{x}^{T}\bm{J}\bm{x}-\bm{x}^{T}(\bm{J}-\bm{A}^{T}\boldsymbol{\Sigma}^{-1}\bm{A})\bm{x}^{(j)}-\bm{x}^{T}\bm{A}^{T}\boldsymbol{\Sigma}^{-1}\bm{r}\nonumber \\
 & +\sum_{n=1}^{N}\Bigg(<\lambda_{0\bm{x}^{(j)}[n]}>\sum_{i_{r}=(n-1)M+1}^{nM}\frac{x_{i_{r}}^{2}+x_{i_{r}+MN}^{2}}{\epsilon}\nonumber \\
+ & <\lambda_{1\bm{x}^{(j)}[n]}>(1+a)\sum_{i_{r}=(n-1)M+1}^{nM}\Big(\frac{x_{i_{r}}^{2}+x_{i_{r}+MN}^{2}}{b\text{(\ensuremath{\frac{(x_{i_{r}}^{(j)})^{2}+(x_{i_{r}+MN}^{(j)})^{2}}{b}}+1)}}\Big)\Bigg)\nonumber \\
 & +g_{1}(\bm{x}^{(j)})\nonumber \\
= & \frac{1}{2}\bm{x}^{T}\bm{J}\bm{x}-\bm{x}^{T}(\bm{J}-\bm{A}^{T}\boldsymbol{\Sigma}^{-1}\bm{A})\bm{x}^{(j)}-\bm{x}^{T}\bm{A}^{T}\boldsymbol{\Sigma}^{-1}\bm{r}+\bm{x}^{T}\bm{\Lambda}_{0}^{(j)}\bm{x}\nonumber \\
 & +\bm{x}^{T}\bm{\Lambda}_{1}^{(j)}\odot\bm{W}^{(j)}\bm{x}+g(\bm{x}^{(j)}),\nonumber 
\end{align}
}where  we have $\bm{J}=\lambda_{max}(\bm{A}^{T}\boldsymbol{\Sigma}^{-1}\bm{A})\bm{I}$,
with $\lambda_{max}(\bm{A}^{T}\boldsymbol{\Sigma}^{-1}\bm{A})$
being the largest eigenvalue of $\bm{A}^{T}\boldsymbol{\Sigma}^{-1}\bm{A}$. $\bm{A}\in\mathbb{R}^{2MT\times2MN}$ and $\boldsymbol{\Sigma}\in\mathbb{R}^{2MT\times2MT}$
are given by Eq.  (\ref{eq:expressr}) and Eq.  (\ref{eq:sigexp}),
respectively. 
We also have
\begin{align}
 & <\lambda_{0\bm{x}^{(j)}[n]}>=\frac{(1-q_{s})P_{0}(\bm{x}^{(j)}[n])}{(1-q_{s})P_{0}(\bm{x}^{(j)}[n])+(q_{s})P_{1}(\bm{x}^{(j)}[n])},\label{eq:l0}\\
 & <\lambda_{1\bm{x}^{(j)}[n]}>=\frac{(q_{s})P_{1}(\bm{x}^{(j)}[n])}{(1-q_{s})P_{0}(\bm{x}^{(j)}[n])+(q_{s})P_{1}(\bm{x}^{(j)}[n])},\label{eq:l1}
\end{align}
with {\scriptsize
\begin{align*}
P_{0}(\bm{x}^{(j)}[n])=\frac{1}{(\pi\epsilon)^{M}}\prod_{i_{r}=(n-1)M+1}^{nM}\exp(-\frac{(x_{i_{r}}^{(j)})^{2}+(x_{i_{r}+MN}^{(j)})^{2}}{\epsilon}\big),
\end{align*}
\begin{align*}
P_{1}(\bm{x}^{(j)}[n])=(\frac{a}{\pi b})^{M}\prod_{i_{r}=(n-1)M+1}^{nM}(\frac{(x_{i_{r}}^{(j)})^{2}+(x_{i_{r}+MN}^{(j)})^{2}}{b}+1)^{-(a+1)},
\end{align*}
}
and {\footnotesize{}$\bm{x}^{(j)}[n]=[x^{(j)}_{(n-1)M+1},\ldots,x^{(j)}_{(N+n)M}]^T$}.
$\bm{\Lambda_{0}}^{(j)}\in\mathbb{R}^{2MN}$, $\bm{\Lambda_{1}}^{(j)}\in\mathbb{R}^{2MN}$,
and $\bm{W}^{(j)}\in\mathbb{R}^{2MN}$ are diagonal
matrices with the diagonal elements given in Table \ref{tab:Value-of-parameters},
where $\mathcal{V}(n)=\{i: i=(n-1)M+1,\ldots,nM,MN+(n-1)M+1,\ldots,(N+n)M\}$
is the index set of $\bm{x}$ related to the $n$-th MTC device. 
\begin{table}[htbp]\scriptsize
\centering{}\caption{Values of parameters in (\ref{eq:update}) \label{tab:Value-of-parameters}}
\begin{tabular}{ccc}
\toprule 
Parameters & \multicolumn{2}{c}{Values}\tabularnewline
\midrule
\midrule 
\multirow{2}{*}{$\bm{\Lambda}_{0(i,i)}^{(j)}$} & $i\in\mathcal{V}(n)$ & \multirow{2}{*}{$\frac{<\lambda_{0\bm{x}^{(j)}[n]}>}{\epsilon}$}\tabularnewline
 &  & \tabularnewline
\midrule 
\multirow{2}{*}{$\bm{\Lambda}_{1(i,i)}^{(j)}$} & $i\in\mathcal{V}(n)$ & \multirow{2}{*}{$\frac{(1+a)<\lambda_{1\bm{x}^{(j)}[n]}>}{b}$}\tabularnewline
 &  & \tabularnewline
\midrule 
\multirow{4}{*}{$\bm{W}_{i,i}^{(j)}$} & $i\leq MN$ & \multirow{2}{*}{$1/(\frac{(x_{i}^{(j)}){}^{2}+(x_{i+MN}^{(j)}){}^{2}}{b}+1)$}\tabularnewline
 &  & \tabularnewline
\cmidrule{2-3} \cmidrule{3-3} 
 & $MN<i\leq2MN$  & \multirow{2}{*}{$1/(\frac{(x_{i}^{(j)}){}^{2}+(x_{i-MN}^{(j)}){}^{2}}{b}+1)$}\tabularnewline
 &  & \tabularnewline
\bottomrule
\end{tabular}
\end{table}

By dropping the constant and irrelevant terms in (\ref{eq:surr1}),
the updated rule of $\bm{x}^{(j+1)}$ is given by

\begin{align}
 & \bm{x}^{(j+1)}\label{eq:update}\\
= & \arg\min\frac{1}{2}\bm{x}^{T}\bm{J}\bm{x}-\bm{x}^{T}(\bm{J}-\bm{A}^{T}\boldsymbol{\Sigma}^{-1}\bm{A})\bm{x}^{(j)}\nonumber \\
& -\bm{x}^{T}\bm{A}^{T}\boldsymbol{\Sigma}^{-1}\bm{r}+\bm{x}^{T}\boldsymbol{\Lambda}_{0}^{(j)}\bm{x}+\bm{x}^{T}\boldsymbol{\Lambda}_{1}^{(j)}\odot\bm{W}^{(j)}\bm{x}\nonumber \\
= & [(\bm{J}-\bm{A}^{T}\boldsymbol{\Sigma}^{-1}\bm{A})\bm{x}^{(j)}+\bm{A}^{T}\boldsymbol{\Sigma}^{-1}\bm{r}]\nonumber \\
 & \times\big(\bm{J}+2\boldsymbol{\Lambda}_{0}^{(j)}+2\boldsymbol{\Lambda}_{1}^{(j)}\odot\bm{W}^{(j)}\big)^{-1}. \nonumber 
\end{align}
Algorithm \ref{alg1} summarizes the key steps of the overall solution
to recover the aggregated variable  from Problem (\ref{eq:prob1}). Due to the page limit, we omit the convergence analysis. However, the convergence of the sequence (\ref{eq:update}) to the stationary point of Problem (\ref{eq:prob1}) can be established easily  by checking the requirements of MM surrogate \cite{sun2017majorization}.
\addtolength{\topmargin}{0.01in}

\begin{algorithm}[htbp]
\small
  \begin{algorithmic}[1]
\State \textbf{Input: $\bm{r}$} , measurement matrix $\boldsymbol{\Phi}$,
and noise variance $\sigma_{v}^{2}.$
\State \textbf{Output:} $\bm{\hat{x}}$
\State \textbf{Initialize: $\bm{x}^{(0)}=\bm{0}$}, $j=0$.
\State \textbf{Pre-computing:}
\State \textbf{\% Calculate coefficients in terms of Bussgang decomposition}
\State  1: Calculate $\bm{R_{n_{q}n_{q}}}$, $\bm{K}$.
\State  2: $\bm{A}=\bm{K}\boldsymbol{\Phi},$ $\boldsymbol{\Sigma}=\bm{\sigma}_{v}^{2}\bm{KK}^{T}+\bm{R}_{\bm{n}_{q}\bm{n}_{q}},$
$\boldsymbol{\Omega}_{1}=\bm{A}^{T}\boldsymbol{\Sigma}^{-1}$,
$\boldsymbol{\Omega}_{2}=\bm{A}^{T}\boldsymbol{\Sigma}^{-1}\bm{A},$
$\bm{f}=\boldsymbol{\Omega}_{1}\bm{r},$ $\bm{J}=\lambda_{max}(\boldsymbol{\Omega}_{2})\bm{I}$.
\State \textbf{Iteration Estimation:}
\State \textbf{\% MM-based solver for estimation of }$\bm{x}$
\State \textbf{while }not converge \textbf{do}
\State  3: Calculate $\bm{\Lambda_{0}}^{(j)}$, $\bm{\Lambda_{1}}^{(j)}$
and $\bm{W}^{(j)}$ via (\ref{eq:l0}), (\ref{eq:l1})
and Table \ref{tab:Value-of-parameters}.
\State  4: Calculate  $\bm{x}^{(j+1)}=[(\bm{J}-\boldsymbol{\Omega}_{2})\bm{x}^{(j)}+\bm{f}](\bm{J}+2\boldsymbol{\Lambda}_{0}^{(j)}+2\boldsymbol{\Lambda}_{1}^{(j)}\odot\bm{W}^{(j)})^{-1}.$\label{-3:line3}
\State 5: $j=j+1$.
\State \textbf{end while }
\State 6:\textbf{ $\bm{\hat{x}}=\bm{x}^{(j+1)}$}.
  \end{algorithmic}
  \caption{\small The Proposed MM-based Algorithm \label{alg1}}
\end{algorithm}

\subsection{Activity Detection and Channel Estimation}

After we obtain $\bm{\hat{x}}$ by Algorithm \ref{alg1}, the MAP estimator of the device activity $s_{n}$ is given by

\begin{equation}
\hat{s}_{n}=\left\{ \begin{array}{rl}
1, & -(1+a)\sum_{i_{r}=(n-1)M+1}^{nM}(\log(\frac{\hat{x}_{i_{r}}^{2}+\hat{x}_{i_{r}+MN}^{2}}{b}+1)\\
 & +\frac{\hat{x}_{i_{r}}^{2}+\hat{x}_{i_{r}+MN}^{2}}{\epsilon})>\varpi_{th},\\
\\
0, & \text{otherwise,}
\end{array}\right.\label{eq:llrde}
\end{equation}
with $\varpi_{th}=-\log q_{s}-M\log(\frac{a}{\pi b})+\log(1-q_{s})+M\log\frac{1}{(\pi\epsilon)}.$

Finally, we have $\hat{\bm{S}}=diag([\hat{s}_{1},\ldots,\hat{s}_{N}]^T)$
and $\hat{\bm{X}}$ can be obtained by reformatting $\bm{\hat{x}}$.

\subsection{Complexity Analysis}
Algorithm \ref{alg1} works for any general pilot. The computation
cost is dominated by the calculation of the inverse of matrix $\boldsymbol{\Sigma}$
and the maximum eigenvalue of $\boldsymbol{\Omega}_{2}$. However, if
the pilot is chosen such that $\bm{D}^{H}\bm{D}$ is diagonal
or the diagonal elements dominate the off-diagonal ones, $\boldsymbol{\Sigma}$
and $\boldsymbol{\Omega}_{2}$ are diagonal or can be well-approximated
by diagonal elements.\footnote{All pilot matrices with sub-Gaussian property can easily satisfy this condition.} In this case,  the dominant complexity of Algorithm \ref{alg1}
can be reduced to $\mathcal{O}((NM)^{2})$. This is because the calculation
in the matrix inversion and eigenvalue can be dramatically reduced by the diagonal approximation
of $\boldsymbol{\Sigma}$ and the most complicated computation is
the matrix-vector product at each iteration (line 4 in Algorithm \ref{alg1}).
In addition, the proposed algorithm only
involves the calculations of  elementary functions in each iteration.
\section{Numerical Simulation and Results Discussion\label{sec:Numerical-Simulation-and}}
In this section, we verify the performance of the proposed joint
AD and CE method. We consider a scenario where $N=200$ online MTC
devices are uniformly distributed in a single BS cell with radius
$R=1$ km, and the device active ratio is $q_{s}=0.1$. The large-scale
fading of the $n$-th MTC device is $10^{-12.81-3.67log_{10}(c_{n})}$,
with $c_{n}$ denoting the distance between the BS and the $n$-th
 device. 
The channel is generated by the widely used LTE spatial channel model
(SCM) \cite{salo2005matlab} with half-wavelength space ULA and $M=128$.
The SNR in the simulations is  the received SNR averaged over
all devices, and the quantization scheme is the optimum quantizer
given in \cite{max1960quantizing}. The AD and CE performance of
the proposed method is compared with  AMP method\cite{chen2018sparse}, Hybrid GAMP method \cite{rangan2012hybrid}, and  GTurbo method \cite{liu2019generalized}.

The performance metric for CE is the estimation mean squared error
(MSE), which is defined as

\begin{equation}
\text{MSE}\triangleq\frac{||\bm{\hat{x}-\bm{x}}||_{2}^{2}}{2NM}.
\end{equation}

To evaluate the AD performance, we compare the true positive rate
(TPR) $\triangleq\frac{\sum_{n=1}^{N}\bm{1}_{(\hat{s}_{n}=1|s_{n}=1)}}{\sum_{n=1}^{N}\bm{1}_{(s_{n}=1)}}$
, false negative rate (FNR) $\triangleq\frac{\sum_{n=1}^{N}\bm{1}_{(\hat{s}_{n}=0|s_{n}=1)}}{\sum_{n=1}^{N}\bm{1}_{(s_{n}=1)}}$,
and false positive rate (FPR) $\triangleq\frac{\sum_{n=1}^{N}\bm{1}_{(\hat{s}_{n}=1|s_{n}=0)}}{\sum_{n=1}^{N}\bm{1}_{(s_{n}=0)}}$
, which can be regarded as the probability of successful detection,
missing detection, and false detection, respectively \cite{nagarajan2014consistent}.

\subsection{Impact of Pilot Length}

First, we compare the performance of different approaches under  different pilot
lengths in Fig.  \ref{fig:DiffT}. Specifically, Fig.  \ref{fig:e2e} shows the  MSE and TPR performance, Fig.  \ref{fig:opl} shows the  FNR and FPR performance. As shown in the results, the proposed method can achieve better performance with fewer pilots compared with the baselines. This is because the proposed method incorporates joint  activity sparsity and angular channel sparsity. 
\begin{figure}
	\centering
	\subfigure[Left: MSE versus pilot length. Right: TPR versus pilot length.]{
		\begin{minipage}[t]{0.98\linewidth}
			\centering
			\includegraphics[width=0.98\linewidth]{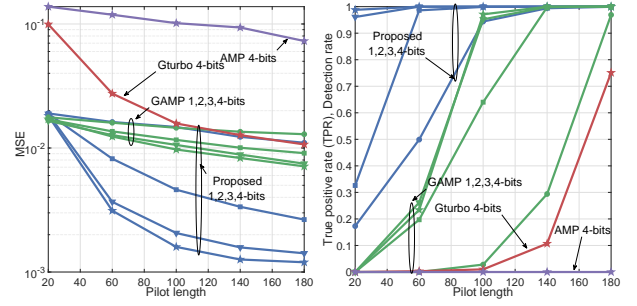}
			\label{fig:e2e}
			%\caption{}
		\end{minipage}%
	}%

	\subfigure[Left: FNR versus pilot length. Right: FPR versus pilot length.]{
		\begin{minipage}[t]{0.98\linewidth}
			\centering
			\includegraphics[width=0.98\linewidth]{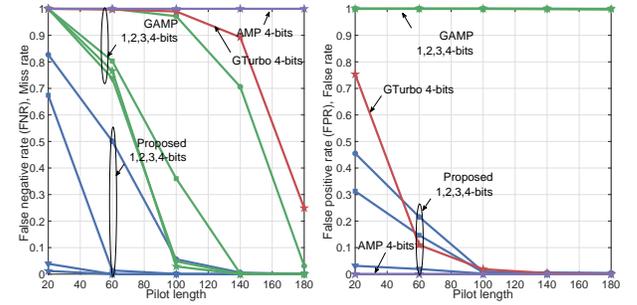}
			\label{fig:opl}
			%\caption{}
		\end{minipage}%
	}
	\centering
	\caption{Performance comparison with different pilot lengths. Set SNR=10 dB, and the pilot sequence is i.i.d. QPSK.\label{fig:DiffT}}
	
\end{figure}

\subsection{Impact of SNR}

We also compare the performance  under different
SNRs in Fig.  \ref{fig:DiffSNR}. The results show that the quantization
effect is significant when the SNR is higher ($>5$dB). In the low
SNR region, the resolution of the ADC
will not affect the performance, while in the high SNR region,
the proposed method can achieve better performance with low-resolution
ADC.

\begin{figure}
	\centering
	\subfigure[Left: MSE versus SNR. Right: TPR versus SNR.]{
		\begin{minipage}[t]{0.98\linewidth}
			\centering
			\includegraphics[width=0.98\linewidth]{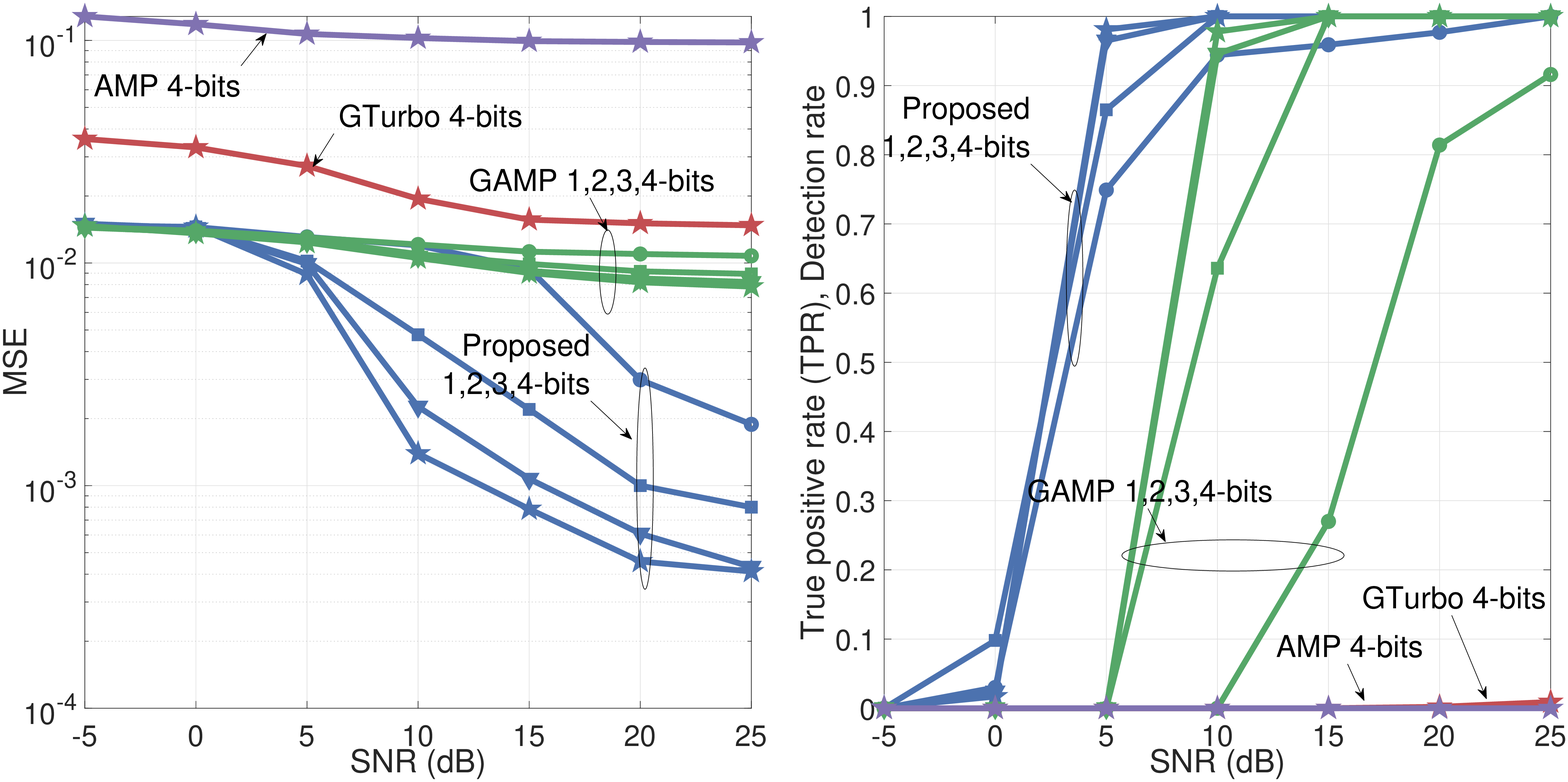}
			\label{fig:e2e}
			%\caption{}
		\end{minipage}%
	}%

	\subfigure[ Left: FNR versus SNR. Right: FPR versus SNR]{
		\begin{minipage}[t]{0.98\linewidth}
			\centering
			\includegraphics[width=0.98\linewidth]{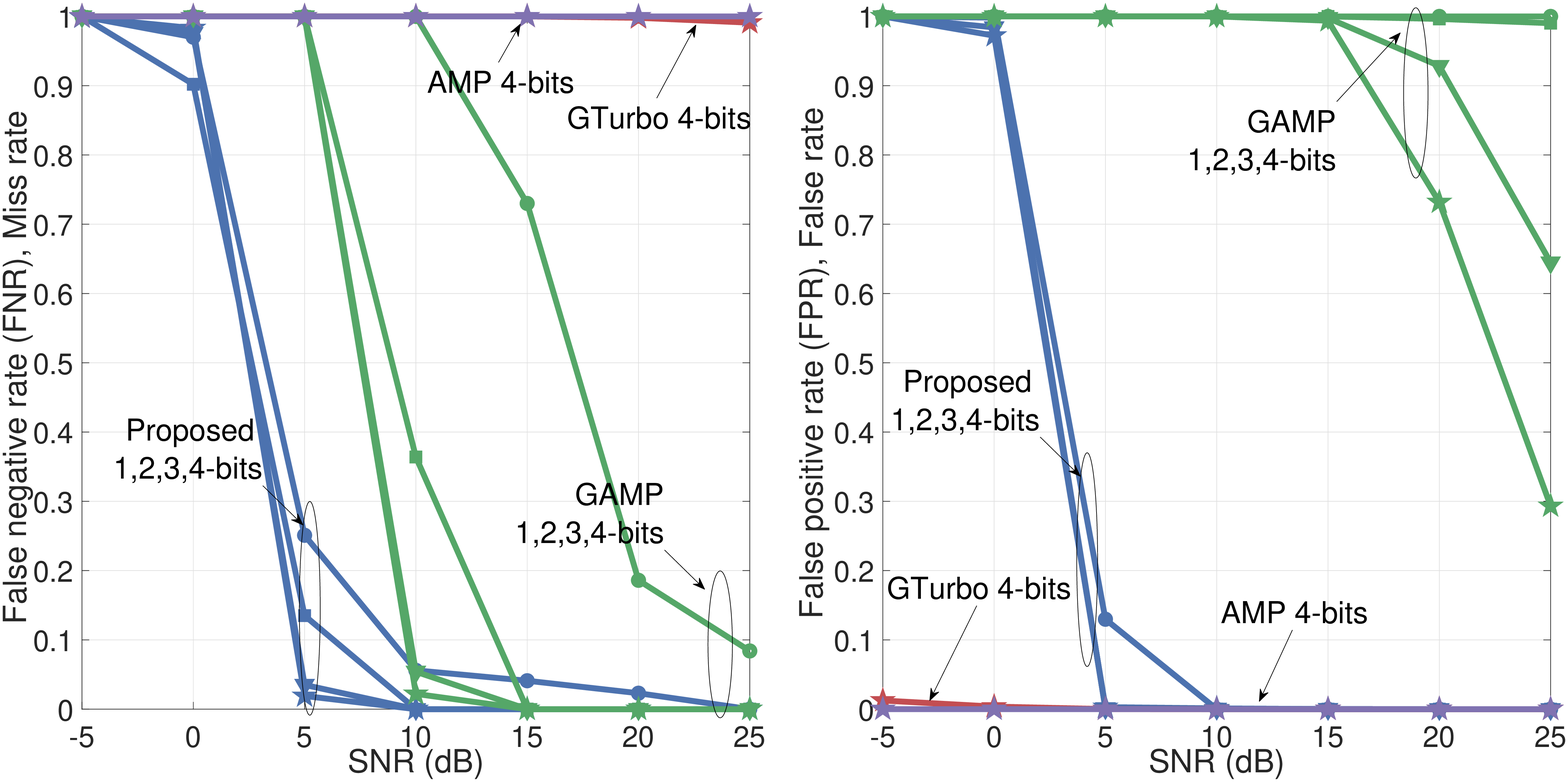}
			\label{fig:opl}
			%\caption{}
		\end{minipage}%
	}
	\centering
	\caption{Performance comparison with different SNRs. Here the pilot sequence is i.i.d. QPSK of the length $T=100$.\label{fig:DiffSNR}}
	
\end{figure}

\section{Conclusions\label{sec:Conclusions}}

In this paper, we  propose a new scheme for the mMTC system  with low-resolution ADC to effectively
detect the active  devices and estimate the corresponding channels.
By incorporating the joint device activity and angular channel sparsity,
an MAP optimization problem is formed and an efficient MM-based algorithm
is given to solve the MAP problem. The proposed scheme has no limitation
to any specific pilot.
Our simulation results demonstrate that  the
proposed method outperforms state-of-the-art schemes.

\bibliographystyle{ieeetr}
\bibliography{myref}

\end{document}